\documentclass[journal,comsoc]{IEEEtran}
\usepackage[dvips]{graphicx}
\usepackage{amsmath}
\usepackage{amssymb}
\usepackage{citesort}
\usepackage{graphicx}
\usepackage{times}
\usepackage{amsfonts}
\usepackage{float}
\usepackage{amsthm}
\usepackage{color}
\usepackage{psfrag}

\newtheorem{algorithm}{Algorithm}

\def\be { \begin{IEEEeqnarray}{rCl} }
\def\ee { \end{IEEEeqnarray} }
\newtheorem{proposition}{Proposition}

\begin{document}

\title{
\huge{Massive MIMO Pilot Retransmission Strategies for Robustification against Jamming}}
\author{
Tan Tai Do, Hien Quoc Ngo, Trung Q. Duong, Tobias J. Oechtering, and Mikael Skoglund\vspace{-0.9cm}\\
\thanks{T. T. Do, H. Q. Ngo is with Link\"oping University, Sweden (e-mail: tan.tai.do@liu.se, nqhien@isy.liu.se). T. J. Oechtering, M. Skoglund are with KTH Royal Institute of Technology, Sweden (e-mail: \{oech,skoglund\}@kth.se).
T. Q. Duong is with  Queen's University Belfast, UK (e-mail: trung.q.duong@qub.ac.uk).}
}
\maketitle

\begin{abstract}
This letter proposes anti-jamming strategies based on pilot
retransmission for a single user uplink massive MIMO under jamming
attack. A jammer is assumed to attack the system both in the
training and data transmission phases. We first derive an
achievable rate which enables us to analyze the effect of jamming
attacks on the system performance. Counter-attack strategies are
then proposed to mitigate this effect under two different
scenarios: random and deterministic jamming attacks. Numerical
results illustrate our analysis and benefit of the proposed
schemes.
\end{abstract}
\vspace{-0.5cm}
\section{Introduction}
\vspace{-0.1cm}
As an emerging candidate for 5G wireless communication networks,
massive multiple-input multiple-output (MIMO) \cite{Mar10TWC,NLM13TCOM}
has drawn a lot of research interests recently. However, there are only
a few works on physical layer security in this area
\cite{Zhu14TWC,ZXu16CL,ZSB16TWC,RZR15CM,BKA15CNS,PRB16WCL,WLW15TWC}.
Among the very few, only some of them have studied jamming aspects although
jamming exists and has been identified as a critical problem for reliable
communications, especially in massive MIMO systems, which are sensitive
to pilot contamination \cite{Mar10TWC}. For instance, the authors consider
security transmission for a downlink massive MIMO system with presence
of attackers capable of jamming and eavesdropping in \cite{RZR15CM,BKA15CNS}.
The problem of smart jamming is considered for an uplink massive MIMO system
in \cite{PRB16WCL}, which shows that a smart jammer can cause pilot
contamination that substantially degrades the system performance.

Most of the above works have been considered from a jammer point
of view: study the jamming strategy, which is the most harmful for
the legitimate user or for the eavesdropper. In this work, we
motivate our study from the system perspective, in which we
develop counter strategies to minimize the effect of jamming
attacks. To this end, we first derive an achievable rate of a
single user uplink massive MIMO with the presence of a jammer.
Then, by exploiting asymptotic properties of massive MIMO systems,
we propose two anti-jamming strategies based on pilot
retransmission protocols for the cases of random jamming and
deterministic jamming attacks. Numerical results show that the
proposed anti-jamming strategies can significantly improve the
system performance.
\vspace{-0.3cm}
\section{Problem Setup}
\begin{figure}[t]
\centering
\psfrag{BS}[][][0.7]{$\mathrm{Base~Station}$}
\psfrag{US}[][][0.7]{$\mathrm{User}$}
\psfrag{JM}[][][0.7]{$\mathrm{Jammer}$}
\psfrag{gu}[][][0.9]{$\mathbf{g}_{\mathrm{u}}$}
\psfrag{gj}[][][0.9]{$\mathbf{g}_{\mathrm{j}}$}
\includegraphics[width=3.5cm]{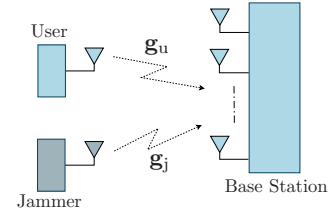}
\caption{Massive MIMO uplink with jamming
attack.\vspace{-0.4cm}}\label{fig:system_model}\vspace{-0.3cm}
\end{figure}

\vspace{-0.1cm}
We consider a single user massive MIMO uplink with the presence of a
jammer as depicted in Fig.~\ref{fig:system_model}. Further, we assume
that the base station (BS) has $M$ antennas, the legitimate user
and the jammer have a single antenna.

Let us denote $\mathbf{g}_\mathrm{u}\in \mathbb{C}^{M\times 1}$ and
$\mathbf{g}_\mathrm{j}\in \mathbb{C}^{M\times 1}$ as the channel
vectors from the user and the jammer to the BS, respectively. We
assume that the elements of $\mathbf{g}_\mathrm{u}$ are
independent and identically distributed (i.i.d.)
zero mean circularly symmetric complex Gaussian (ZMCSCG) random
variables, i.e., $\mathbf{g}_\mathrm{u}\sim
\mathcal{CN}(0,\beta_\mathrm{u}\mathbf{I}_M)$, where
$\beta_\mathrm{u}$ represents the large-scale fading (path loss
and shadowing). Similarly, we assume that
$\mathbf{g}_\mathrm{j}\sim
\mathcal{CN}(0,\beta_\mathrm{j}\mathbf{I}_M)$ and  is independent
of $\mathbf{g}_\mathrm{u}$.

We consider a block-fading model, in which the channel remains
constant during a coherence block of $T$ symbols, and varies
independently from one coherence block to the
next. To communicate with the BS, the legitimate user follows a
two-phase TDD transmission protocol: i) Phase 1: the user sends
pilot sequences to the BS for channel estimation, and
ii) Phase 2: the user transmits the payload data to the BS.
We assume that the jammer attacks the uplink transmission both in
the training and in the data payload transmission phases.
\vspace{-0.4cm}

\subsection{Training phase}
During the first $\tau$ channel uses ($\tau<T$), the user sends to
the BS a pilot sequence $\sqrt{\tau p_t}\mathbf{s}_\mathrm{u}$,
where $p_t$ is the transmit pilot power and
$\mathbf{s}_{\mathrm{u}}\in \mathbb{C}^{\tau \times 1}$
originates from a pilot codebook $\mathcal{S}$ containing $\tau$
orthogonal unit-power vectors. At the same time, the jammer sends
$\sqrt{\tau q_t}\mathbf{s}_\mathrm{j}$ to interfere the channel
estimation, where $\mathbf{s}_{\mathrm{j}}\in \mathbb{C}^{\tau
\times 1}$ satisfies
$\mathbb{E}\{\|\mathbf{s}_{\mathrm{j}}\|^2\}=1$ and $q_t$ is the
transmit power of the jammer during the training phase.
Accordingly, the received signals at the BS is given by \be
\label{Ytm} \mathbf{Y}_t=\sqrt{\tau
p_t}\mathbf{g}_\mathrm{u}\mathbf{s}_\mathrm{u}^T+\sqrt{\tau
q_t}\mathbf{g}_\mathrm{j}\mathbf{s}_\mathrm{j}^T+\mathbf{N}_t, \ee
where $\mathbf{N}_t \in \mathbb{C}^{M \times \tau}$ is the
additive noise matrix with unit power i.i.d. ZMCSCG elements.

The BS then performs a de-spreading operation as:
\be \label{yt}
\mathbf{y}_t=\mathbf{Y}_t\mathbf{s}_\mathrm{u}^*
    =
    \sqrt{\tau
p_t}\mathbf{g}_\mathrm{u}+\sqrt{\tau
q_t}\mathbf{g}_\mathrm{j}\mathbf{s}_\mathrm{j}^T
\mathbf{s}_\mathrm{u}^* + \tilde{\mathbf{n}}_t, \ee where
$\tilde{\mathbf{n}}_t \triangleq
\mathbf{N}_t\mathbf{s}_\mathrm{u}^*$ and $\tilde{\mathbf{n}}_t \sim
\mathcal{CN}(0,\mathbf{I}_{M})$. The minimum mean squared error
(MMSE) estimate of $\mathbf{g}_\mathrm{u}$ given $\mathbf{y}_t$ is
\cite{Kay93} \be \label{guh}
\mathbf{\hat{g}}_\mathrm{u}=c_\mathrm{u}\mathbf{y}_t, \ee where
$c_\mathrm{u}=\frac{\sqrt{\tau p_t}\beta_\mathrm{u}}{\tau
p_t\beta_\mathrm{u}+\tau
q_t\beta_\mathrm{j}|\mathbf{s}_\mathrm{j}^T\mathbf{s}_\mathrm{u}^*|^2+1}$.

The MMSE estimator (\ref{guh}) requires that the BS has to know
$\beta_\mathrm{u}$, $\beta_\mathrm{j}$, and
$|\mathbf{s}_\mathrm{j}^T\mathbf{s}_\mathrm{u}^*|$. Since
$\beta_\mathrm{u}$ and $\beta_\mathrm{j}$ are large-scale fading
coefficients which change very slowly with time (some $40$ times
slower than the small-scale fading coefficients), they can be
estimated at the BS easily \cite{AML:IT:2014}. The quantity
$|\mathbf{s}_\mathrm{j}^T\mathbf{s}_\mathrm{u}^*|$ includes the
jamming sequence $\mathbf{s}_\mathrm{j}$ which is unknown at
the BS. However, by exploiting asymptotic properties of the
massive MIMO, the BS can estimate
$|\mathbf{s}_\mathrm{j}^T\mathbf{s}_\mathrm{u}^*|$ from the
received pilot signal $\mathbf{Y}_t$. We will discuss about this
in detail in Section~\ref{sec:MP}.

Let
$\mathbf{e}_{\mathbf{u}}=\mathbf{g_\mathrm{u}}-\mathbf{\hat{g}}_\mathrm{u}$
be the channel estimation error. From the properties of MMSE
estimation, $\mathbf{\hat{g}}_\mathrm{u}$ and
$\mathbf{e}_{\mathbf{u}}$ are independent. Furthermore, we have
$\mathbf{\hat{g}}_\mathrm{u}\sim\mathcal{CN}(0,\gamma_{\mathrm{u}}\mathbf{I}_{M})$
and
$\mathbf{e}_{\mathbf{u}}\sim\mathcal{CN}(0,(\beta_\mathrm{u}-
\gamma_{\mathrm{u}})\mathbf{I}_{M})$, where
\be \label{gamu} \gamma_{\mathrm{u}}\triangleq\frac{\tau
p_t\beta_\mathrm{u}^2}{\tau p_t\beta_\mathrm{u}+\tau
q_t\beta_\mathrm{j}|\mathbf{s}_\mathrm{j}^T\mathbf{s}_\mathrm{u}^*|^2+1}=c_\mathrm{u}\sqrt{\tau
p_t}\beta_\mathrm{u}. \ee
\vspace{-0.5cm}

\subsection{Data transmission phase}
During the last $(T-\tau)$ channel uses, the user transmits the
payload data to the BS and the jammer continues to interfere with
its jamming signal. Let  $x_\mathrm{u}$
($\mathbb{E}\{\|x_\mathrm{u}\|^2\}=1$) and $x_\mathrm{j}$
($\mathbb{E}\{\|x_\mathrm{j}\|^2\}=1$) be the transmitted signals
from the user and the jammer, respectively. The BS receives
\be
\mathbf{y}_d=\sqrt{p_d}\mathbf{g}_\mathrm{u}x_\mathrm{u}+
\sqrt{q_d}\mathbf{g}_\mathrm{j}x_\mathrm{j}+\mathbf{n}_d,
\ee
where $p_d$ and $q_d$ are the transmit powers from the user
and jammer in the data transmission phase, respectively. The noise
vector  $\mathbf{n}_d$ is assumed to have i.i.d.
$\mathcal{CN}(0,1)$ elements.

To estimate $x_\mathrm{u}$, the BS performs the maximal ratio
combining based on the estimated channel $\mathbf{\hat{g}}_\mathrm{u}$
as follow
\be
\label{y}
y=\mathbf{\hat{g}}_\mathrm{u}^H\mathbf{y}_d
=\sqrt{p_d}\mathbf{\hat{g}}_\mathrm{u}^H\mathbf{g}_\mathrm{u}x_\mathrm{u}+
\sqrt{q_d}\mathbf{\hat{g}}_\mathrm{u}^H\mathbf{g}_\mathrm{j}x_\mathrm{j}+
\mathbf{\hat{g}}_\mathrm{u}^H\mathbf{n}_d.
\ee
\vspace{-0.7cm}
\section{Achievable Rate and Impact of Jamming Attack} \label{sec:ratenjam}
In order to analyze the impact of jamming attack on the system, we
derive a capacity lower bound (achievable rate) for the massive
MIMO channel, described as in (\ref{y}). Substituting
$\mathbf{g}_\mathrm{u}=\mathbf{\hat{g}}_\mathrm{u}+\mathbf{e}_\mathrm{u}$ into (\ref{y}),
we have
\be \label{yb}
y\!=\!\sqrt{p_d}\|\mathbf{\hat{g}}_\mathrm{u}\|^2x_\mathrm{u}+
\!\sqrt{p_d}\mathbf{\hat{g}}_\mathrm{u}^H\!\mathbf{e}_\mathrm{u}x_\mathrm{u}+
\!\sqrt{q_d}\mathbf{\hat{g}}_\mathrm{u}^H\!\mathbf{g}_\mathrm{j}x_\mathrm{j}+
\mathbf{\hat{g}}_\mathrm{u}^H\!\mathbf{n}_d.
\ee

Since $y$ consists of the signals associated with the channel uncertainty and
jamming, we derive an achievable rate using the method suggested in \cite{JAM11TWC}.
To this end, we decompose the received signal in (\ref{yb}) as
\be
\label{yc}
&&y=\sqrt{p_d}\mathbb{E}\{\|\mathbf{\hat{g}}_\mathrm{u}\|^2\}x_\mathrm{u}+ \\
\nonumber
&&\underbrace{\sqrt{p_d}(\|\mathbf{\hat{g}}_\mathrm{u}\|^2\!-
\!\mathbb{E}\{\|\mathbf{\hat{g}}_\mathrm{u}\|^2\}\!+
\mathbf{\hat{g}}_\mathrm{u}^H\mathbf{e}_\mathrm{u} )x_\mathrm{u}\!+\! \sqrt{q_d}\mathbf{\hat{g}}_\mathrm{u}^H\mathbf{g}_\mathrm{j}x_\mathrm{j}+
\mathbf{\hat{g}}_\mathrm{u}^H\mathbf{n}_d.}_{\triangleq ~n_{\mathrm{eff}} ~
- ~\mathrm{effective~ noise}}
\ee

Since $n_{\mathrm{eff}}$ and the desired signal are uncorrelated,
we can obtain an achievable rate by treating $n_{\mathrm{eff}}$ as the
worst-case Gaussian noise, which can be characterized as follows.

\begin{proposition}\label{pro}
An achievable rate of the massive MIMO channel with jamming is
\be
\label{R}
R=\left(1-\tau/T\right)\log_2\left(1+\rho\right),
\ee
where $\rho$ is the effective SINR, given by
\be
\label{rho}
\rho=\frac{Mp_d\gamma_\mathrm{u}}{p_d\beta_\mathrm{u}+q_d\beta_\mathrm{j}+
M\frac{q_dq_t}{p_t}\left(\frac{\beta_\mathrm{j}}{\beta_\mathrm{u}}\right)^2
|\mathbf{s}_\mathrm{j}^T\mathbf{s}_\mathrm{u}^*|^2\gamma_\mathrm{u}+1}.
\ee
\begin{proof}
See Appendix~\ref{Pro1Proof}.
\end{proof}
\end{proposition}\vspace{-0.1cm}
Two interesting remarks can be made:
\begin{itemize}
    \item[(i)] If the jammer does not attack during the training phase
($q_t=0$) or $\mathbf{s}_\mathrm{j}$ and $\mathbf{s}_\mathrm{u}$
are orthogonal
($\mathbf{s}_\mathrm{j}^T\mathbf{s}_\mathrm{u}^*=0$), the
achievable rate becomes
    \be
    \label{R_barj}
     R
     &=&
     \left(1-\frac{\tau}{T}\right)\log_2\Big(1+\frac{Mp_d\gamma_\mathrm{u}}
     {p_d\beta_\mathrm{u}+q_d\beta_\mathrm{j}+1}\Big)
     \mathop \to \limits^{M\to\infty}\infty.~~
    \vspace{-0.2cm}\ee
The achievable rate increases without bound as $M\to\infty$, even
when the jammer attacks during the data transmission phase.

\item[(ii)] If the jammer attacks during the training phase and
$\mathbf{s}_\mathrm{j}^T\mathbf{s}_\mathrm{u}^*\neq0$, we have
\vspace{-0.4cm}
\be
R  \mathop \to \limits^{M\to\infty}&&
  \left(1-\frac{\tau}{T}\right)\log_2\left(\frac{p_tp_d}{q_tq_d}
\frac{\beta_\mathrm{u}^2}{\beta_\mathrm{j}^2}\frac{1}
{|\mathbf{s}_\mathrm{j}^T\mathbf{s}_\mathrm{u}^*|^2}\right).
\vspace{-0.2cm}\ee
This implies that when the training phase is attacked, the
achievable rate is rapidly saturated even when $M\to\infty$. This
is the effect of \emph{jamming-pilot contamination}.
\end{itemize}

\vspace{-0.25cm}
\section{Pilot Retransmission Scheme}

As discussed in Section~\ref{sec:ratenjam}, the jamming attack
during the training phase highly affects the system performance.
Therefore, we focus on the training phase and construct counter
strategies to mitigate the effect of jamming-pilot
contamination. We propose pilot retransmission schemes where the
pilot will be retransmitted when the jamming-pilot
contamination is high ($|\mathbf{s}_\mathrm{j}^T\mathbf{s}_\mathrm{u}^*|$ is large).
Note that, some overheads for synchronization are necessary for
the pilot retransmission protocols. However, those overheads are
negligible compared to the payload data.

\vspace{-0.3cm}
\subsection{Mathematical Preliminaries}\label{sec:MP}
We show that by exploiting asymptotic properties
of the massive MIMO system, the BS can estimate
$|\mathbf{s}_\mathrm{j}^T\mathbf{s}_\mathrm{u}^*|$ and
$\mathbf{s}_\mathrm{j}^*\mathbf{s}_\mathrm{j}^T$ from the received
pilot signals $\mathbf{y}_t$ and $\mathbf{Y}_t$ even
$\mathbf{s}_\mathrm{j}$ is unknown.
\subsubsection{Estimation of $|\mathbf{s}_\mathrm{j}^T\mathbf{s}_\mathrm{u}^*|^2$}
By the law of large numbers,
    \be\label{eq:as1}
    \frac{1}{M}\|\mathbf{y}_t\|^2
    &&= \tau p_t\frac{\|\mathbf{g}_\mathrm{u}\|^2}{M} + \tau
    q_t|\mathbf{s}_\mathrm{j}^T\mathbf{s}_\mathrm{u}^*|^2
    \frac{\|\mathbf{g}_\mathrm{j}\|^2}{M}+ \frac{\|\tilde{\mathbf{n}}_t\|^2}{M}
    \nonumber\\
    &&+ \sqrt{\tau p_t}\frac{\mathbf{g}_\mathrm{u}^H
    \left(\sqrt{\tau q_t}\mathbf{g}_\mathrm{j}\mathbf{s}_\mathrm{j}^T \mathbf{s}_\mathrm{u}^*
     + \tilde{\mathbf{n}}_t\right)}{M}\nonumber\\
     &&+\sqrt{\tau q_t}\mathbf{s}_\mathrm{u}^T \mathbf{s}_\mathrm{j}^*
    \frac{\mathbf{g}_\mathrm{j}^H \left(\sqrt{\tau p_t}\mathbf{g}_\mathrm{u}+
    \tilde{\mathbf{n}}_t\right)}{M}
    \nonumber\\
    &&+ \frac{\tilde{\mathbf{n}}_t^H \left(\sqrt{\tau p_t}\mathbf{g}_\mathrm{u}+
    \sqrt{\tau q_t}\mathbf{g}_\mathrm{j}\mathbf{s}_\mathrm{j}^T \mathbf{s}_\mathrm{u}^*
    \right)}{M}\nonumber\\
    &&
    \mathop \to \limits^{a.s.} \tau p_t\beta_\mathrm{u}+\tau q_t|\mathbf{s}_\mathrm{j}^T
    \mathbf{s}_\mathrm{u}^*|^2\beta_\mathrm{j} + 1, ~\mathrm{as}
    ~M\to\infty,\vspace{-0.4cm}\ee
where $\mathop\to \limits^{a.s.}$ denotes almost sure convergence. From
\eqref{eq:as1}, and under the assumption that the BS knows
$\beta_\mathrm{u}$ and $\beta_\mathrm{j}$,
$|\mathbf{s}_\mathrm{j}^T\mathbf{s}_\mathrm{u}^*|^2$ can be
estimated as
\vspace{-0.2cm}
\be\label{eq:as2}
    \widehat{|\mathbf{s}_\mathrm{j}^T\mathbf{s}_\mathrm{u}^*|^2}
    =    \frac{1}{\tau q_t M\beta_\mathrm{j}}\|\mathbf{y}_t\|^2
    -    \frac{ p_t \beta_\mathrm{u} }{q_t\beta_\mathrm{j}}-
    \frac{1}{\tau q_t\beta_\mathrm{j}}.
\vspace{-0.1cm}
\ee
\subsubsection{Estimation of
$\mathbf{s}_\mathrm{j}^*\mathbf{s}_\mathrm{j}^T$}\label{sec:estjj}
From \eqref{Ytm} and again from the law of large numbers, as
$M\to\infty$, we have
\be
    \frac{1}{M}\mathbf{Y}_t^H\mathbf{Y}_t
    \mathop \to \limits^{a.s.}
    \tau p_t\beta_{\mathrm{u}}\mathbf{s}_\mathrm{u}^*\mathbf{s}_\mathrm{u}^T
    +\tau q_t\beta_{\mathrm{j}}\mathbf{s}_\mathrm{j}^*\mathbf{s}_\mathrm{j}^T
    + \mathbf{I}_\tau.
\ee

Thus, the BS can estimate
$\mathbf{s}_\mathrm{j}^*\mathbf{s}_\mathrm{j}^T$ as
\be\label{eq:es11}
    \widehat{\mathbf{s}_\mathrm{j}^*\mathbf{s}_\mathrm{j}^T}
    = \frac{1}{\tau q_t\beta_{\mathrm{j}}M}\mathbf{Y}_t^H\mathbf{Y}_t
    -\frac{ p_t\beta_{\mathrm{u}}}{ q_t\beta_{\mathrm{j}}}
     \mathbf{s}_\mathrm{u}^*\mathbf{s}_\mathrm{u}^T
    - \frac{1}{\tau q_t\beta_{\mathrm{j}}}\mathbf{I}_\tau.
\ee
Based on the estimates of
$|\mathbf{s}_\mathrm{j}^T\mathbf{s}_\mathrm{u}^*|^2$ and
$\mathbf{s}_\mathrm{j}^*\mathbf{s}_\mathrm{j}^T$, in next
sections, we propose two pilot retransmission schemes to deal with
two common jamming cases: random and deterministic
jamming.

\vspace{-0.5cm}
\subsection{Pilot Retransmission under Random Jamming}\label{sect:rj}
In practice, if the jammer does not have the prior knowledge of the pilot
sequences used by the user, then it will send a random sequence to attack
the system. During the training phase, the user sends a pilot sequence
$\mathbf{s}_\mathrm{u}\in \mathcal{S}$, while
the  jammer sends a random jamming sequence. The BS estimates
$|\mathbf{s}_\mathrm{j}^T\mathbf{s}_\mathrm{u}^*|^2$ and requests
the user to retransmit a new pilot sequence until
$|\mathbf{s}_\mathrm{j}^T\mathbf{s}_\mathrm{u}^*|^2$ is smaller than a
threshold $\varepsilon$ or the number of transmissions exceeds
the maximum number $N_{\mathrm{max}}$. The pilot retransmission
algorithm is summarized as follows:

\begin{algorithm}[Under random jamming]\label{sec:algo1}
~
\vspace{-0.05cm}\begin{itemize}
   \item[1.] Initialization: set $N=1$, choose the values of pilot
   length $\tau$, threshold $\varepsilon$, and
   $N_{\mathrm{max}}$ ($N_{\mathrm{max}}\tau< T$).

   \item[2.] User sends a random $\tau\times 1$ pilot sequence
   $\mathbf{s}_\mathrm{u}\in \mathcal{S}$.

   \item[3.] The BS estimates
   $|\mathbf{s}_\mathrm{j}^T\mathbf{s}_\mathrm{u}^*|^2$ using
   \eqref{eq:as2}. If $|\mathbf{s}_\mathrm{j}^T\mathbf{s}_\mathrm{u}^*|^2
   \leq \varepsilon$ or $N=N_{\mathrm{max}}$ $\rightarrow$ Stop.
   Otherwise, go to step 4.

   \item[4.] Set $N = N+1$,  go to step 2.
\end{itemize}
\end{algorithm}

Let $\mathbf{s}_\mathrm{u}(n)$ and $\mathbf{s}_\mathrm{j}(n)$ be
the pilot and jamming sequences respectively, corresponding to the $n$th
retransmission,  $n=1,...,N$. Similar to \eqref{R}, the achievable rate
of the massive MIMO with anti-jamming for random jamming is given
by
\be \label{R_rj}
R_{\mathrm{rj}}=\left(\!1\!-\!\frac{N\tau}{T}\!\right)\!\log_2\!
\left(\!1+\frac{Mp_d\gamma_\mathrm{u}}{p_d\beta_\mathrm{u}+q_d\beta_\mathrm{j}
+\alpha_{\mathrm{rj}}+1}\!\right),
\ee
where $\alpha_{\mathrm{rj}}=M\frac{q_dq_t}{p_t}\frac{\beta_\mathrm{j}^2}{\beta_\mathrm{u}^2}
\min\limits_{n}|\mathbf{s}_\mathrm{j}(n)^T\mathbf{s}_\mathrm{u}(n)^*|^2\gamma_\mathrm{u}$.
Note that in order to realize the achievable rate
$R_{\mathrm{rj}}$ in (\ref{R_rj}), the BS has to buffer the
received pilot signal then processes with the best one (with
minimal $|\mathbf{s}_\mathrm{j}(n)^T\mathbf{s}_\mathrm{u}(n)^*|^2$)
after $N$ pilot retransmissions.
There exists case where $|\mathbf{s}_\mathrm{j}(1)^T\mathbf{s}_\mathrm{u}(1)^*|^2$ is
minimum which degrades the system performance since it consumes more training resource
without finding a better candidate. However, the pilot is only retransmitted when the
first transmission is bad, and hence, there will be a high probability that the
retransmission is better than the first one.

\vspace{-0.5cm}
\subsection{Pilot Retransmission under Deterministic Jamming}\label{sec:dj}
Next, we assume that the jamming sequences are deterministic
during the training phase, i.e.,
$\mathbf{s}_\mathrm{j}(1)=\cdots=\mathbf{s}_\mathrm{j}(N)$. Such
scenario can happen, for instance, in case the jammer has the
prior knowledge of the pilot length and pilot sequence codebook and tries to attack
using a deterministic function of those training sequences \cite{BKA15CNS}. In this
case, the massive MIMO system can outsmart the jammer by adapting the training
sequences based on the knowledge on the current pilot transmission instead of just
randomly retransmitting them as in the previous case.

We observe that
$|\mathbf{s}_\mathrm{j}^T\mathbf{s}_\mathrm{u}^*|^2$ can be
decomposed as
\be
|\mathbf{s}_\mathrm{j}^T\mathbf{s}_\mathrm{u}^*|^2=\mathbf{s}_\mathrm{u}^T
\mathbf{s}_\mathrm{j}^*\mathbf{s}_\mathrm{j}^T\mathbf{s}_\mathrm{u}^*.
\ee
So, if the BS knows $\mathbf{s}_\mathrm{j}^*\mathbf{s}_\mathrm{j}^T$, it can
choose $\mathbf{s}_\mathrm{u}$ to minimize $|\mathbf{s}_\mathrm{j}^T\mathbf{s}_\mathrm{u}^*|$. In
Section~\ref{sec:estjj}, we know that the BS can estimate
$\mathbf{s}_\mathrm{j}^*\mathbf{s}_\mathrm{j}^T$ from $\mathbf{Y}_t$. From this
observation, we propose the following pilot retransmission scheme:

\begin{algorithm}[Under deterministic jamming]\label{sec:
algo2}
~
\vspace{-0.05cm}\begin{itemize}
   \item[1.] Initialization: choose the values of pilot length $\tau$ and threshold $\varepsilon$.

   \item[2.] User sends a $\tau\times 1$ pilot sequence $\mathbf{s}_\mathrm{u}\in \mathcal{S}$.

   \item[3.] The BS estimates $|\mathbf{s}_\mathrm{j}^T\mathbf{s}_\mathrm{u}^*|^2$ using
   \eqref{eq:as2}. If $|\mathbf{s}_\mathrm{j}^T\mathbf{s}_\mathrm{u}^*|^2 \leq
   \varepsilon$ $\rightarrow$ Stop. Otherwise, go to step 4.

   \item[4.] The BS  estimates $\mathbf{s}_\mathrm{j}^*\mathbf{s}_\mathrm{j}^T$ using
   \eqref{eq:es11}. Then, the BS finds $\mathbf{s}_\mathrm{u}^{\mathrm{opt}}$ so that
   $\mathbf{s}_\mathrm{u}^{\mathrm{opt}T}\mathbf{s}_\mathrm{j}^*
\mathbf{s}_\mathrm{j}^T\mathbf{s}_\mathrm{u}^{\mathrm{opt}*}$ is
   minimal. If $\mathbf{s}_\mathrm{u}^{\mathrm{opt}T}\mathbf{s}_\mathrm{j}^*
  \mathbf{s}_\mathrm{j}^T\mathbf{s}_\mathrm{u}^{\mathrm{opt}*} <
 |\mathbf{s}_\mathrm{j}^T\mathbf{s}_\mathrm{u}^*|^2$, then the user will retransmit
  this new pilot.
\end{itemize}
\end{algorithm}

Since the BS requests the user to retransmit its pilot only if
$|\mathbf{s}_\mathrm{j}^T\mathbf{s}_\mathrm{u}^*|$ of the first
transmission exceeds the threshold $\varepsilon$, the achievable
rate is \be \label{R_dj}
R_{\mathrm{dj}}\!=\!\left(\!1\!-\!\frac{N\tau}{T}\!\right)\!\log_2\!\Big(1\!+
\!\frac{Mp_d\gamma_\mathrm{u}}{p_d\beta_\mathrm{u}+q_d\beta_\mathrm{j}+
\alpha_{\mathrm{dj}}+1}\!\Big), \ee where $\left\{\!\!\!\!
\begin{array}{l}
  \alpha_{\mathrm{dj}}=M\frac{q_dq_t}{p_t}\left(\frac{\beta_\mathrm{j}}
  {\beta_\mathrm{u}}\right)^2|\mathbf{s}_\mathrm{j}^T\mathbf{s}_\mathrm{u}^*|^2
  \gamma_\mathrm{u}, N=1, ~ \text{if} ~
|\mathbf{s}_\mathrm{j}^T\mathbf{s}_\mathrm{u}^*|
\leq \varepsilon \\
  \alpha_{\mathrm{dj}}=M\frac{q_dq_t}{p_t}\left(\frac{\beta_\mathrm{j}}
  {\beta_\mathrm{u}}\right)^2|\mathbf{s}_\mathrm{j}^T
  \mathbf{s}_\mathrm{u}^{\mathrm{opt}*}|^2\gamma_\mathrm{u}, N=2, ~
\text{otherwise}. \\
\end{array}%
\right.$ The maximal number of retransmissions for this case is
one.

\section{Numerical Results}
In this section, we numerically evaluate the performance of the
proposed anti-jamming schemes in term of the average achievable rate.
The average is taken over 50000 realizations of $(\mathbf{s}_\mathrm{u},\mathbf{s}_\mathrm{j})$.
We assume that the transmit powers at the user and jammer satisfy $\tau p_t+(T-\tau)p_d\leq TP$
and $\tau q_t+(T-\tau)q_d\leq TQ$. We also assume $T=200$ channel uses, maximum number of
transmissions $N_{\mathrm{max}}=2$.

\begin{figure}[t]
\centering
\includegraphics[width=7.5cm]{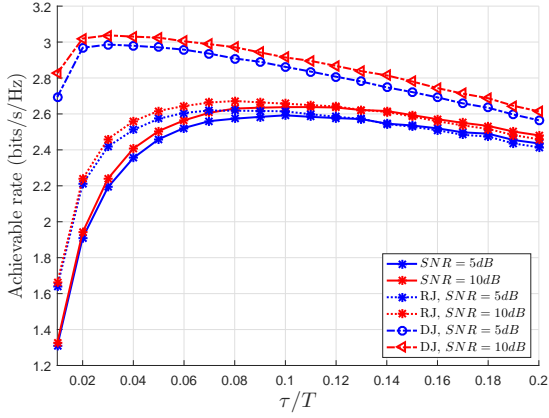}
\caption{Average achievable rates of different anti-jamming schemes for
$\varepsilon=0.1$, $q_t=p_t=q_d=p_d=SNR$, $M=50$. The solid curves, dotted curves (with label ``RJ''),
and dashed curves (with label ``DJ'') denote the achievable rates without pilot retransmission
(c.f. Proposition 1), with counter strategy for random jamming (c.f. Alg. 1), and with counter
strategy for deterministic jamming (c.f. Alg. 2), respectively.}
\label{vstau}
\end{figure}

Fig. \ref{vstau} illustrates the average achievable rates for different anti-jamming schemes
according to the training payload ($\tau/T$). It shows that in order to achieve the best
performances, the training payloads should be selected properly to balance the channel estimation
quality ($\tau$ is large enough) and the resource allocated for data transmission ($\tau$ is not
too large). As expected, the proposed schemes with pilot retransmission outperform the conventional
scheme (without pilot retransmission). When the training sequence is very long, i.e., $\tau/T$ is
large, the proposed schemes are close to the conventional one since the probability of pilot
retransmission is very small as the channel estimation quality is often good enough after the
first training transmission. Note that in this simulation, we choose $\varepsilon=0.1$ which
is not optimal in general. It is expected that the benefits of the our proposed schemes are even
larger with optimal $\varepsilon$.

\begin{figure}[t]
\centering
\includegraphics[width=7.5cm]{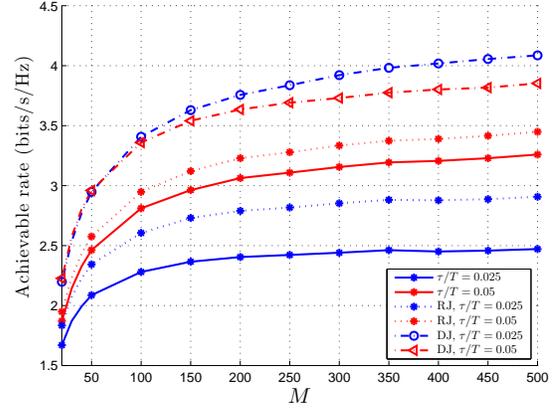}
\caption{Average achievable rates according to the number of antennas $M$ for $\varepsilon=0.1$,
$p_t=q_t=p_d=q_d=SNR=5dB$. The legend for curves is similar to Fig. \ref{vstau}.}
\label{vsM}
\end{figure}

Figure~\ref{vsM} shows the average achievable rates versus the number of BS
antennas. Without anti-jamming strategy, the pilot contamination
can severely harm the system performance and obstruct the scaling
of achievable rate with $M$. This is consistent with our analysis
in Section III. The performance can be remarkably improved by using
the pilot retransmission protocols. Particularly, for the case of
deterministic jamming, the proposed scheme can overcome the pilot
contamination bottleneck and allows the achievable rates scale with
$M$ even when $M$ is large.
\vspace{-0.1cm}
\section{Conclusion}
\vspace{-0.1cm}
The problem of anti-jamming for a single-user uplink massive MIMO
has been considered. It showed that jamming attacks could severely
degrade the system performance. By exploiting the asymptotic
properties of large antenna array, we proposed two pilot
retransmission protocols. With our proposed schemes, the pilot
sequences and training payload could flexibly be adjusted to
reduce the effect of jamming attack and improve the system
performance.

Future work may study multi-user networks. For instance, our results can be readily
extended if a max-min fairness criterion is used. Then the pilot retransmission protocol
design is considering the worst user who has the smallest achievable rate.

\vspace{-0.2cm}
\appendices
\section{Proof of Proposition 1}\label{Pro1Proof}
\vspace{-0.2cm}
By treating $n_{\mathrm{eff}}$ as Gaussian additive noise, an achievable rate of the
channel in (\ref{yc}) is given by
 \be\label{eq:proofrate1a}
R&=&     \left(1-\frac{\tau}{T}\right)\log_2\left(1+\frac{p_d|\mathbb{E}\{\|\mathbf{\hat{g}}_\mathrm{u}\|^2\}|^2}
{\mathbb{E}\{|n_{\mathrm{eff}}|^2\}}\right).
\ee
Let us define
 \be\label{eq:proofrate1}
\rho\triangleq\frac{p_d|\mathbb{E}\{\|\mathbf{\hat{g}}_\mathrm{u}\|^2\}|^2}
{\mathbb{E}\{|n_{\mathrm{eff}}|^2\}}\triangleq\frac{p_d
        M^2\gamma_\mathrm{u}^2}{E_1+E_2+E_3},
\ee
where $E_1\triangleq p_d\mathbb{E}\left\{\left|\|\mathbf{\hat{g}}_\mathrm{u}\|^2-
\mathbb{E}\left\{\|\mathbf{\hat{g}}_\mathrm{u}\|^2\right\}+
\mathbf{\hat{g}}_\mathrm{u}^H\mathbf{e}_\mathrm{u}\right|^2
\right\}$, $E_2\triangleq
q_d\mathbb{E}\left\{\left|\mathbf{\hat{g}}_\mathrm{u}^H\mathbf{g}_\mathrm{j}\right|^2\right\}$,
and $E_3\triangleq
\mathbb{E}\left\{|\mathbf{\hat{g}}_\mathrm{u}^H\mathbf{n}_d|^2\right\}$.
Since $\mathbf{\hat{g}}_\mathrm{u}$ and $\mathbf{e}_\mathrm{u}$
are independent zero mean random vectors, we have \be\label{eq:E1}
E_1
    &=&
p_d\mathbb{E}\{\|\mathbf{\hat{g}}_\mathrm{u}\|^4\}-p_d(\mathbb{E}\{\|\mathbf{\hat{g}}_\mathrm{u}\|^2\})^2 + q_d\mathbb{E}\{|\mathbf{\hat{g}}_\mathrm{u}^H\mathbf{e}_\mathrm{u}|^2\}\nonumber\\
    &=&
    p_dM(M+1)\gamma_\mathrm{u}^2-p_dM^2\gamma_\mathrm{u}^2+
p_dM\gamma_\mathrm{u}(\beta_\mathrm{u}-\gamma_\mathrm{u})\nonumber\\
 &=&M\gamma_\mathrm{u}p_d\beta_\mathrm{u}.
\ee

From (\ref{guh}), and using the fact that
$\mathbf{g}_\mathrm{u},\mathbf{g}_\mathrm{j},
\tilde{\mathbf{n}}_t$ are independent and zero mean random
vectors, we have
 \be
\nonumber
E_2
    &=&
    q_dc_\mathrm{u}^2\mathbb{E}\left\{\left|\sqrt{\tau
p_t}\mathbf{g}_\mathrm{u}^H\mathbf{g}_\mathrm{j}+ \sqrt{\tau
q_t}\|\mathbf{g}_\mathrm{j}\|^2\mathbf{s}_\mathrm{u}^T\mathbf{s}_\mathrm{j}^*+
\tilde{\mathbf{n}}_t^H\mathbf{g}_\mathrm{j}\right|^2\right\}\nonumber\\
    &\overset{(a)}{=}&
    q_dc_\mathrm{u}^2\left(\tau
p_t\mathbb{E}\!\left\{\left|\mathbf{{g}}_\mathrm{u}^H\mathbf{g}_\mathrm{j}\right|^2\right\}\!+\!
\tau
q_t|\mathbf{s}_\mathrm{u}^T\mathbf{s}_\mathrm{j}^*|^2\mathbb{E}\{\|\mathbf{g}_\mathrm{j}\|^4\}\!
    \!+\!\mathbb{E}\!\left\{\left|\tilde{\mathbf{n}}_t^H\mathbf{g}_\mathrm{j}\right|^2\right\}\right)
\nonumber\\
\nonumber
    &=&
    q_dc_\mathrm{u}^2(\tau p_tM\beta_\mathrm{u}\beta_\mathrm{j}+\tau q_tM(M+1)\beta_\mathrm{j}^2|\mathbf{s}_\mathrm{u}^T\mathbf{s}_\mathrm{j}^*|^2+M\beta_\mathrm{j}).\ee
Then by using (\ref{gamu}),
 \be\label{eq:E2}
E_2
=
    M q_d\gamma_u\left(\beta_\mathrm{j}+M\gamma_u\frac{q_t}{p_t}\frac{\beta_\mathrm{j}^2}{\beta_\mathrm{u}^2}
|\mathbf{s}_\mathrm{j}^T\mathbf{s}_\mathrm{u}^*|^2\right). \ee
Similarly,
\be\label{eq:E3}
E_3=\mathbb{E}\{|\mathbf{\hat{g}}_\mathrm{u}^H\mathbf{n}_d|^2\}=M\gamma_\mathrm{u}.
\ee

Substituting \eqref{eq:E1}, \eqref{eq:E2}, and \eqref{eq:E3} into
\eqref{eq:proofrate1} we obtain \eqref{R}.


\begin{thebibliography}{10}
\providecommand{\url}[1]{#1}
\csname url@samestyle\endcsname
\providecommand{\newblock}{\relax}
\providecommand{\bibinfo}[2]{#2}
\providecommand{\BIBentrySTDinterwordspacing}{\spaceskip=0pt\relax}
\providecommand{\BIBentryALTinterwordstretchfactor}{4}
\providecommand{\BIBentryALTinterwordspacing}{\spaceskip=\fontdimen2\font plus
\BIBentryALTinterwordstretchfactor\fontdimen3\font minus
  \fontdimen4\font\relax}
\providecommand{\BIBforeignlanguage}[2]{{%
\expandafter\ifx\csname l@#1\endcsname\relax
\typeout{** WARNING: IEEEtran.bst: No hyphenation pattern has been}%
\typeout{** loaded for the language `#1'. Using the pattern for}%
\typeout{** the default language instead.}%
\else
\language=\csname l@#1\endcsname
\fi
#2}}
\providecommand{\BIBdecl}{\relax}
\BIBdecl

\bibitem{Mar10TWC}
T.~L. Marzetta, ``Noncooperative cellular wireless with unlimited numbers of
  base station antennas,'' \emph{IEEE Trans. on Wireless Commun.}, vol.~9,
  no.~11, pp. 3590--3600, Nov. 2010.

\bibitem{NLM13TCOM}
H.~Q. Ngo, E.~G. Larsson, and T.~L. Marzetta, ``Energy and spectral efficiency
  of very large multiuser {MIMO} systems,'' \emph{{IEEE} Trans. Commun.},
  vol.~61, no.~4, pp. 1436--1449, Apr. 2013.

\bibitem{Zhu14TWC}
J.~Zhu, R.~Schober, and V.~K. Bhargava, ``Secure transmission in multicell
  massive mimo systems,'' \emph{IEEE Trans. on Wireless Commun.}, vol.~13,
  no.~9, pp. 4766--4781, Sep. 2014.

\bibitem{ZXu16CL}
J.~Zhu and W.~Xu, ``Securing massive {MIMO} via power scaling,'' \emph{IEEE
  Commun. Letters}, vol.~20, no.~5, pp. 1014--1017, 2016.

\bibitem{ZSB16TWC}
J.~Zhu, R.~Schober, and V.~K. Bhargava, ``Linear precoding of data and
  artificial noise in secure massive {MIMO} systems,'' \emph{IEEE Trans. on
  Wireless Commun.}, vol.~15, no.~3, pp. 2245--2261, 2016.

\bibitem{RZR15CM}
D.~Kapetanovic, G.~Zheng, and F.~Rusek, ``Physical layer security for massive
  {MIMO}: An overview on passive eavesdropping and active attacks,'' \emph{IEEE
  Commun. Maga.}, vol.~53, no.~6, pp. 21--27, 2015.

\bibitem{BKA15CNS}
Y.~O. Basciftci, C.~E. Koksal, and A.~Ashikhmin, ``Securing massive {MIMO} at
  the physical layer,'' in \emph{IEEE Conf. on Commun. and Net. Sec. (CNS)
  2015}, Philadelphia, PA, USA, Sep. 2015, pp. 272--280.

\bibitem{PRB16WCL}
H.~Pirzadeh, S.~M. Razavizadeh, and E.~Bj{\"{o}}rnson, ``Subverting massive
  {MIMO} by smart jamming,'' \emph{IEEE Wireless Commun. Letters}, vol.~5,
  no.~1, pp. 20--23, Feb. 2016.

\bibitem{WLW15TWC}
J.~Wang, J.~Lee, F.~Wang, and T.~Q.~S. Quek, ``Jamming-aided secure
  communication in massive {MIMO} {Rician} channels,'' \emph{IEEE Trans. on
  Wireless Commun.}, vol.~14, no.~12, pp. 6854--6868, 2015.

\bibitem{Kay93}
S.~M. Kay, \emph{Fundamentals of Statistical Signal Processing: Estimation
  Theory}.\hskip 1em plus 0.5em minus 0.4em\relax NJ, USA: Prentice-Hall, 1993.

\bibitem{AML:IT:2014}
A.~Ashikhmin, T.~L. Marzetta, and L.~Li, ``Interference reduction in multi-cell
  massive {MIMO} systems i: Large-scale fading precoding and decoding,''
  submitted to \emph{IEEE Trans. Inf. Theory} 2014.

\bibitem{JAM11TWC}
J.~Jose, A.~Ashikhmin, T.~L. Marzetta, and S.~Vishwanath, ``Pilot contamination
  and precoding in multi-cell {TDD} systems,'' \emph{IEEE Trans. on Wireless
  Commun.}, vol.~10, no.~8, pp. 2640--2651, Aug. 2011.

\end{thebibliography}
\end{document}